\documentclass[aps,pra,noshowpacs,twocolumn,preprintnumbers,superscriptaddress,floatfix]{revtex4}
\usepackage{amsmath,amssymb}
\usepackage{multirow}
\usepackage{hyperref,graphicx}
\usepackage{epsfig}
\usepackage{times}
\usepackage{pdfpages}
\usepackage[Symbol]{upgreek}

\begin{document}
\title{Mapping and Measuring Large-scale Photonic Correlation with Single-photon Imaging}

\author{Ke Sun}
\affiliation{State Key Laboratory of Advanced Optical Communication Systems and Networks, School of Physics and Astronomy, Shanghai Jiao Tong University, Shanghai 200240, China}
\author{Jun Gao}
\affiliation{State Key Laboratory of Advanced Optical Communication Systems and Networks, School of Physics and Astronomy, Shanghai Jiao Tong University, Shanghai 200240, China}
\affiliation{Synergetic Innovation Center of Quantum Information and Quantum Physics, University of Science and Technology of China, Hefei, Anhui 230026, China}
\author{Ming-Ming Cao}
\affiliation{State Key Laboratory of Advanced Optical Communication Systems and Networks, School of Physics and Astronomy, Shanghai Jiao Tong University, Shanghai 200240, China}
\author{Zhi-Qiang Jiao}
\affiliation{State Key Laboratory of Advanced Optical Communication Systems and Networks, School of Physics and Astronomy, Shanghai Jiao Tong University, Shanghai 200240, China}
\affiliation{Synergetic Innovation Center of Quantum Information and Quantum Physics, University of Science and Technology of China, Hefei, Anhui 230026, China}
\author{Yu Liu}
\affiliation{State Key Laboratory of Advanced Optical Communication Systems and Networks, School of Physics and Astronomy, Shanghai Jiao Tong University, Shanghai 200240, China}
\author{Zhan-Ming Li}
\affiliation{State Key Laboratory of Advanced Optical Communication Systems and Networks, School of Physics and Astronomy, Shanghai Jiao Tong University, Shanghai 200240, China}
\affiliation{Synergetic Innovation Center of Quantum Information and Quantum Physics, University of Science and Technology of China, Hefei, Anhui 230026, China}
\author{Eilon Poem}
\affiliation{Clarendon Laboratory, University of Oxford, Parks Road, Oxford OX1 3PU, UK}
\affiliation{Department of Physics of Complex Systems, Weizmann Institute of Science, Rehovot 7610001, Israel}
\author{Andreas Eckstein}
\affiliation{Clarendon Laboratory, University of Oxford, Parks Road, Oxford OX1 3PU, UK}
\author{Ruo-Jing Ren}
\affiliation{State Key Laboratory of Advanced Optical Communication Systems and Networks, School of Physics and Astronomy, Shanghai Jiao Tong University, Shanghai 200240, China}
\affiliation{Synergetic Innovation Center of Quantum Information and Quantum Physics, University of Science and Technology of China, Hefei, Anhui 230026, China}
\author{Xiao-Ling Pang}
\affiliation{State Key Laboratory of Advanced Optical Communication Systems and Networks, School of Physics and Astronomy, Shanghai Jiao Tong University, Shanghai 200240, China}
\affiliation{Synergetic Innovation Center of Quantum Information and Quantum Physics, University of Science and Technology of China, Hefei, Anhui 230026, China}
\author{Hao Tang}
\affiliation{State Key Laboratory of Advanced Optical Communication Systems and Networks, School of Physics and Astronomy, Shanghai Jiao Tong University, Shanghai 200240, China}
\affiliation{Synergetic Innovation Center of Quantum Information and Quantum Physics, University of Science and Technology of China, Hefei, Anhui 230026, China}

\author{Ian A. Walmsley}
\affiliation{Clarendon Laboratory, University of Oxford, Parks Road, Oxford OX1 3PU, UK}

\author{Xian-Min Jin}

\affiliation{State Key Laboratory of Advanced Optical Communication Systems and Networks, School of Physics and Astronomy, Shanghai Jiao Tong University, Shanghai 200240, China}
\affiliation{Synergetic Innovation Center of Quantum Information and Quantum Physics, University of Science and Technology of China, Hefei, Anhui 230026, China}
\thanks{xianmin.jin@sjtu.edu.cn}
\date{\today}
\maketitle

%\section{Abstract}
%\section{Main}
\textbf{Quantum correlation and its measurement are essential in exploring fundamental quantum physics problems and developing quantum enhanced technologies. Quantum correlation may be generated and manipulated in different spaces, which demands different measurement approaches corresponding to position, time, frequency and polarization of quantum particles. In addition, after early proof-of-principle demonstrations, it is of great demand to measure quantum correlation in a Hilbert space large enough for real quantum applications. When the number of modes goes up to several hundreds, it becomes economically unfeasible for single-mode addressing and also extremely challenging for processing correlation events with hardware. Here we present a general and large-scale measurement approach of Correlation on Spatially-mapped Photon-Level Image (COSPLI). The quantum correlations in other spaces are mapped into the position space and are captured by single-photon-sensitive imaging system. Synthetic methods are developed to suppress noises so that single-photon registrations can be faithfully identified in images. We eventually succeed in retrieving all the correlations with big-data technique from tens of millions of images. We demonstrate our COSPLI by measuring the joint spectrum of parametric down-conversion photons. Our approach provides an elegant way to observe the evolution results of large-scale quantum systems, representing an innovative and powerful tool added into the platform for boosting quantum information processing.}

\noindent Quantum correlation, as one of the unique features of quantum theory, plays a crucial role in quantum information applications. After experiencing quantum evolutions, e.g. quantum interference, quantum particles can be correlated in more diverse ways than the classical counterpart. For example, Hong-Ou-Mandel (HOM) interference can reveal the nonclassical bunching properties of photons\cite{Hong1987Measurement}. These correlation characteristics are crucial in quantum computing\cite{Aaronson2011The,Broome2013Photonic,Spring2013Boson,Bentivegna2015Experimental}, quantum simulation\cite{Alberto,Kon,JO,+++OE Gao Jun} and quantum communication\cite{Pan-tele,Jin-tele,QKD-1ST,Seawater}. 

In theory, especially in the field of quantum computing, if we manage to send enough entangled photons into plenty of modes and operate their superposition states simultaneously, we would be able to obtain sufficiently large quantum state space that may enable a higher computational power than classical computers. In practice, it may still be acceptable to place single photon detector behind each spatial mode for comparably small systems \cite{+++OE Gao Jun, Carolan2013On, Spagnolo2013General, Lebugle2015Experimental}. However, it will become both technically challenging and economically unfeasible to address thousands of modes simultaneously with single photon detectors, and therefore become a decisive bottleneck preventing from detecting state spaces large enough for real quantum applications. 

Thankfully, recent advances of charge-coupled device (CCD) cameras make it possible to directly image spatial output results at a single-photon level \cite{1-Science,2-Nature,3-PRB,4-Nano left,++++++nature communication,Poland-Hologram,Poland-NC,Tangeaat3174}. HOM interference was also successfully verified by low-noise correlation detection on two modes with an intensified camera\cite{++++sCMOS}. 

An alternative and elegant approach we present in this work is to convert the modes in different degrees of freedom into the modes in position, and then measure all the spatial modes accordingly by the large number of units in single-photon-sensitive cameras. We call this general and large-scale measurement approach as Correlation on Spatially-mapped Photon-Level Image (COSPLI). We have also developed the methods to retrieve the low-noise signal of single-photon registrations and their correlations from tens of millions of images. As an example for applications, we experimentally demonstrate our COSPLI by measuring the spectral correlations \cite{Eckstein2011,Gerrits2015,Poem2016,Davis2017} of parametric down-conversion photons.

\begin{figure*}[!ht]
	\centering
	\includegraphics[width=0.90\textwidth]{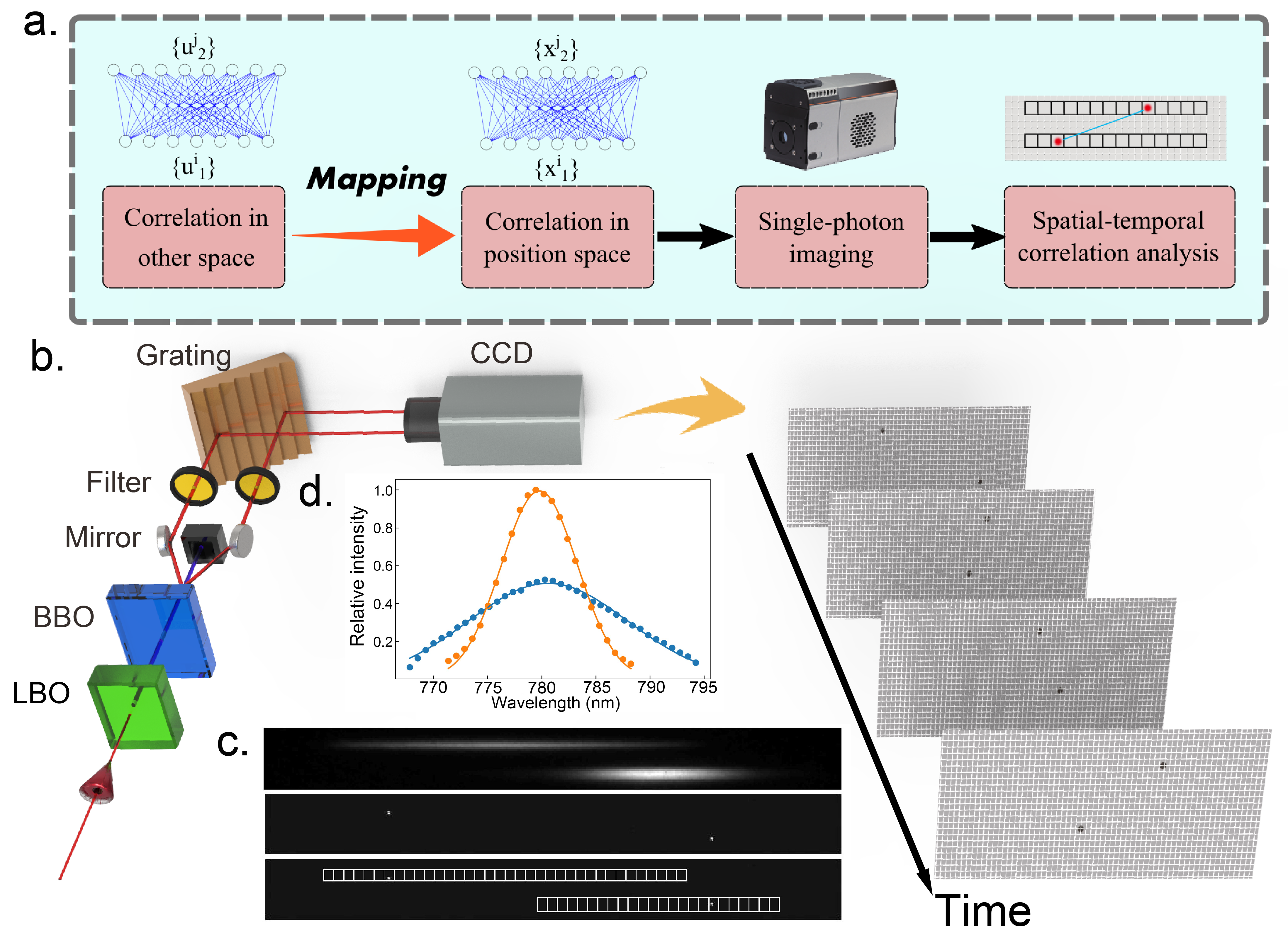}
	\caption{\textbf{a.} Schematic of COSPLI. \textbf{b.}  Experimental setup. A mode-lock Ti:sapphire oscillator, centered at $780nm$ with a laser pulse duration of $130fs$ and a repetition rate of $77$ MHz, pumps a $1.3$-mm thick $LiB_3O_5$(LBO) crystal to double the laser frequency to $390nm$. The ultraviolet laser then pumps a $2$-mm thick $\beta$-$BaB_2O_4$(BBO) crystal to create identical photon pairs via a $\chi^{(2)}$ type-\uppercase\expandafter{\romannumeral2} spontaneous parametric down-conversion process in the beam-like scheme \cite{PhysRevA.68.013804}. The correlation in frequency space of signal and idler photons are mapped into the correlation in position space by a blazed grating with a line density of $1200 l/mm$. \textbf{c. upper} An accumulated image with all spectral components. The upper and lower spots are signal and idler photons, respectively. \textbf{c. middle} One of the frames that possess two registered photons. The two shiny spots are distributed in the two characterized areas in the accumulated image, respectively. \textbf{c. lower} The white grids define the resolution of correlation measurement by dividing the mapped space into many individual parts.
\textbf{d.} The spectra of signal and idler photons derived from the accumulated image.} 
	\label{setup}
\end{figure*}

COSPLI can generally be achieved in five steps (Figure.\ref{setup}\textbf{a}). The first step is to identify the target space where correlation exists. The quantum states of a single photon can be expressed in various eigenvectors with corresponding eigenvalues, such as the position, time, frequency and polarization. The target space can be whichever space needed to measure. In Figure.\ref{setup}\textbf{a}, we use symbol $\{u^i_1\}$ and $\{u^j_2\}$ to represent the target variables. In our experiments, we detect joint spectrum of correlated photons to demonstrate this technique. Under this circumstance, the target is spectrum or frequency space. 

The target variable is usually hard to be directly detected in large number of modes. Therefore, a mapping operation is necessary to convert the target to position space, $\{x^i_1\}$ and $\{x^j_2\}$, which can be measured by a CCD camera. To demonstrate this, we map the frequency information of photons to their positions. In optical system, prisms and gratings are commonly used as spectroscopic devices. To optimize the efficiency and splitting angle, we choose a blazed grating to transform different frequency components of photons into their corresponding positions. Within a particular wavelength range, the first-order diffracted spot of blazed grating is the brightest, rather than the zero-order reflected spot, which means that the grating loss can be controlled to a relatively low level. 

The mapped positions of photons can be either discrete or continuous. For instance, the outputs of two-dimensional quantum walk from a photonic chip present a discrete spot array. Thus, we are able to directly divide them into several discrete regions according to the location of spots, and construct a correlation matrix. As is shown in Figure.\ref{setup}\textbf{a}, symbols $\{x^i_1\}$ and $\{x^j_2\}$ represent different components of converted position variables. In our demonstration experiment, the frequency spectra of photons are continuous, and the converted positions therefore are also continuous. However, in the data analysis process, we have to divide the continuous pattern into equal-size segments. As a result, the continuity of data is destroyed to some extent. Hence, we need to restore its inherent continuity at a later stage.

The forth step is to measure the spatially-mapped positions with a proper imaging system. Nowadays, intensified charge-coupled device (ICCD) and intensified scientific complementary metal oxide semiconductor (IsCMOS) \cite{Angle-momentum} are being well developed to observe single photons. Their high time resolution can be employed as temporal filter to suppress noises. We retrieve the spatial information from each frame, and retrieve the temporal information from the order of frames. We then can perform the last step, which is to calculate the correlation matrix to analyse spatial and/or temporal correlation. 

\begin{figure*}[!ht]
	\centering
	\includegraphics[width=1.0\textwidth]{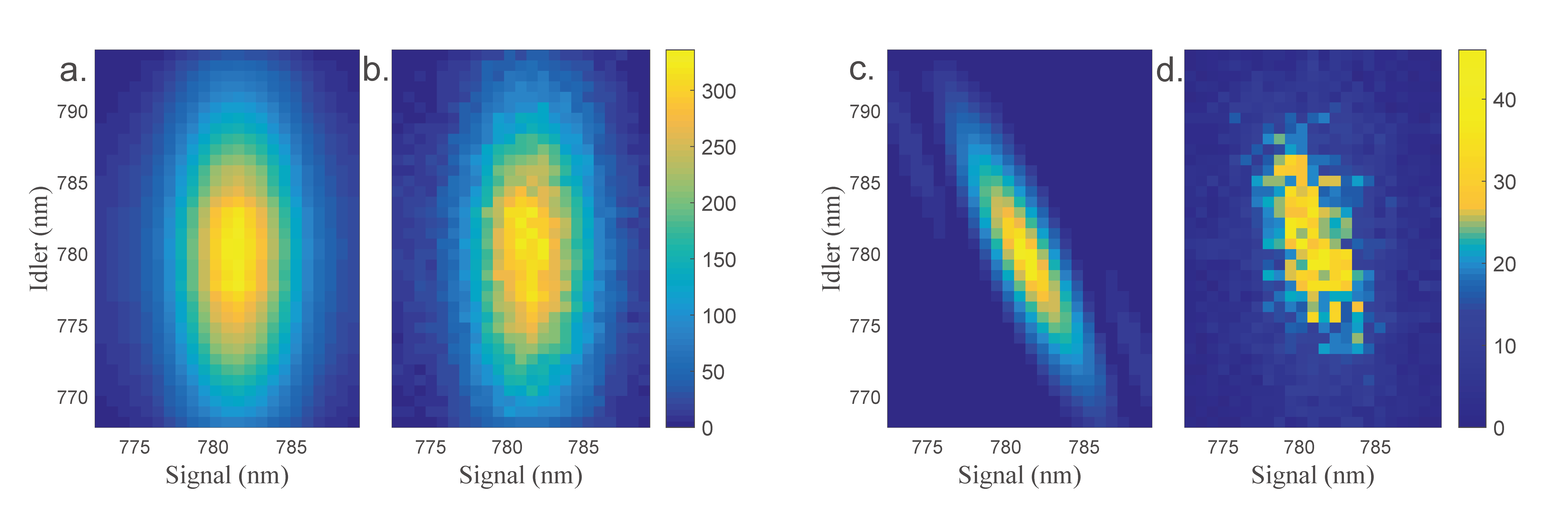}
	\caption{\textbf{Theoretical and experimental results of correlation measurement}. \textbf{a.} and \textbf{b.} are theoretical and experimental results of time-independent joint spectral intensity matrix of photons. The vertically-elliptical shape suggests the photons share no correlation. \textbf{c.} and \textbf{d.} are theoretical and experimental results of time-dependent joint spectral intensity matrix. The joint spectrum of correlated photons generated by BBO is obliquely-elliptical shape rather than a strict one-to-one relationship, because the two photons are not maximally correlated in frequency space\cite{Grice2003Eliminating}.} 
	\label{Figure2}
\end{figure*}  

As an example, we demonstrate a mapping from frequency correlation to spatial correlation of photons (see Figure.\ref{setup} {\bf b}). Type-\uppercase\expandafter{\romannumeral2} spontaneous parametric down-conversion\cite{kwiat1995new} is applied to generate two frequency-correlated photons, the joint spectrum of which can be written as
\begin{equation}
	S_{dep}(\omega_s,\omega_i) = |f(\omega_s,\omega_i)|^2 = |A\alpha(\omega_s+\omega_i)\phi(\omega_s,\omega_i)|^2
	\label{joint S}
\end{equation}(see Methods). The correlated photons emitted by beta-barium-borate (BBO) crystal are delivered to the blazed grating, After mapping their frequency information to position information, signal and idler photons are then captured by an ICCD. 

In the correlation analysis process, we firstly define the regions where photons will appear by accumulating signal and idler photons on a single image. We then divide the regions of signal and idler photons into 24 and 37 segments respectively as shown in Figure.\ref{setup} {\bf c}, with a spatial interval of 10 pixels. Each segment represents one component of the spectrum, and the number of counts of every segment indicates the intensity of each component. In this way, the continuous information in spectral domain is transformed to the discrete spatial positions. Based on the accumulated intensity and the calibrated spectra, we express the spectrum distribution of the two photons as $p_s$ and $p_i$ shown in Figure.\ref{setup} {\bf d}. 

Figure.\ref{setup} {\bf c} shows several shiny points that can be distinguished from the background. Although they may be the result of dark counts, these events are all considered being generated by actual photons. We assume that the noises are generated with a same probability in every pixel of the camera, resulting in a noise matrix $N(\omega_s,\omega_i) = p_{ns}(\omega_s)p_{ni}(\omega_i)$, which means they should just contribute to a contour base of the final correlation matrix as $S(\omega_s,\omega_i) + N(\omega_s,\omega_i)$. With an optimized criterion, we use a computer program to automatically identify the existence of registered photons and record their positions $x_s$, $x_i$, and then input this correlation information into correlation matrix {\it S}($\omega_s$, $\omega_i$) (see Methods).

\begin{figure*}[!th]
	\centering
	\includegraphics[width=1.03\textwidth]{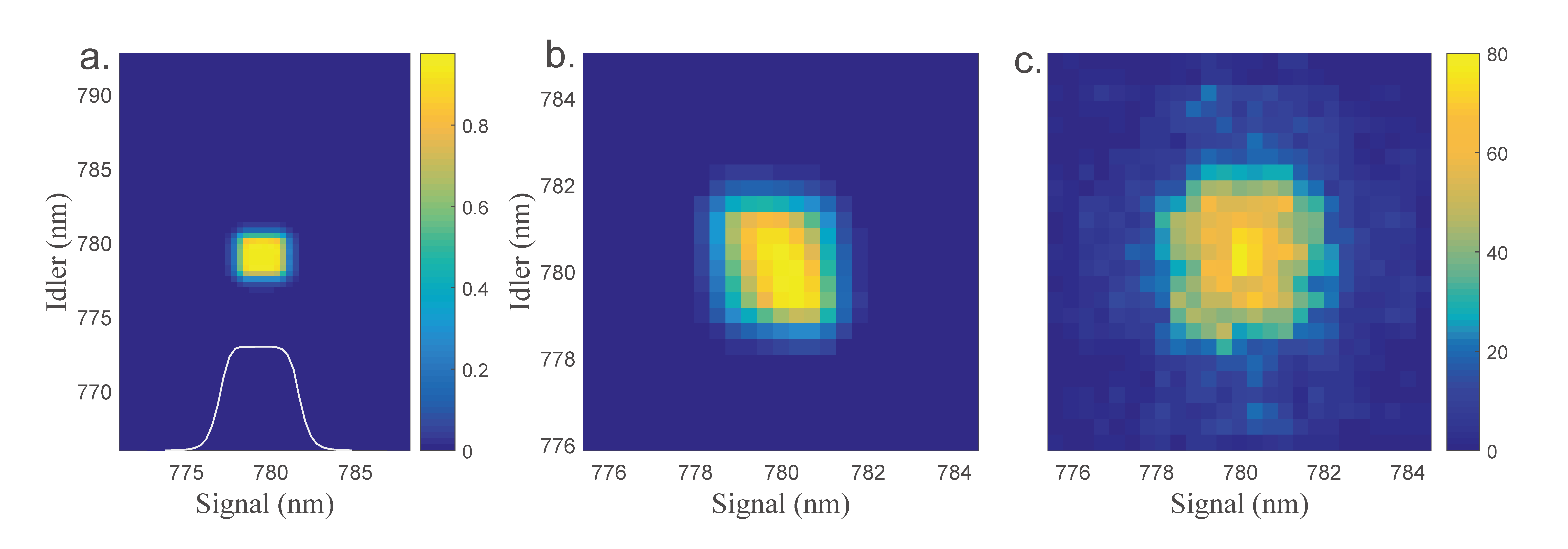}
	\caption{\textbf{Theoretical and experimental results of filtered joint spectrum}. \textbf{a.} The theoretical spectrum reshaped by filters. The white curve is the spectrum transmission rate of the bandpass filter, centered at $779.5nm$ and with a full width at half maximum (FWHM) of $3nm$. \textbf{b.} and \textbf{c.} are theoretical and experimental result of joint spectral intensity matrix of correlated photons.} 
	\label{3nm}
\end{figure*} 
 
COSPLI is a technique that allows the acquisition of spatial-temporal correlation of photons. In our experiments, both time-independent and time-dependent spatial correlations are demonstrated. The time width of the camera gate determines whether photons in one frame share correlation or not. In the time-independent scenario, we set ICCD gate width as $10 \mu s$, far longer than the pump laser pulse interval ($12.9 ns$). Hundreds of photon pairs may appear in one frame, however, on account of many factors of loss, only one or two photons finally register in one frame. In this case, photons detected in the same frame are seldom originally generated by the same pump pulse, which means they are not temporally correlated, and therefore have no frequency correlation. The statistic characteristic of these photons will be the same as that of an ensemble, where particles are all independent. The correlation matrix turns out to be simply the product of probability distribution
\begin{equation}
	S_{indep}(\omega_s,\omega_i) = p_s(\omega_s)p_i(\omega_i),
\end{equation} 
where $p_s(\omega_s)$ and $p_i(\omega_i)$ represent the spectrum distribution of signal and idler photons. $S_{indep}$ is the joint spectrum intensity matrix, representing the probability that signal and idler photon simultaneously possess frequency $\omega_s$ and $\omega_i$, respectively \cite{Grice2003Eliminating}. Unlike the joint spectrum of correlated photons (Eqn.\ref{joint S}), the time-independent one is separable. Figure.\ref{Figure2}\textbf{b} shows the experimental results obtained from 100,000 frames. The perfect similarity with the prediction (Figure.\ref{Figure2}\textbf{a}) implies that most of shiny points on the screen are real photons rather than dark counts or environmental noise, which can serve as a test of single-photon sensitivity of ICCD. 

Now we consider the time-dependent scenario, in which only the two correlated photons generated by the same pump pulse appear in the same frame. We synchronize the generation and detection of correlated photons by triggering ICCD, with an electric pulse produced by laser oscillator. Another main difference with the time-independent scenario is the gate width. For the purpose of ensuring that every frame contains at most one pair of the correlated photons, the gate width is set as $12.5 ns$, which is shorter than the interval between two laser pulses ($12.9 ns$). In this case, detected photons should share the expected frequency correlation with each other. We put the frames that contain correlation events together in a video (see Supplementary Material). 

\begin{figure}[htb!]
	\centering
	\includegraphics[width=0.5\textwidth]{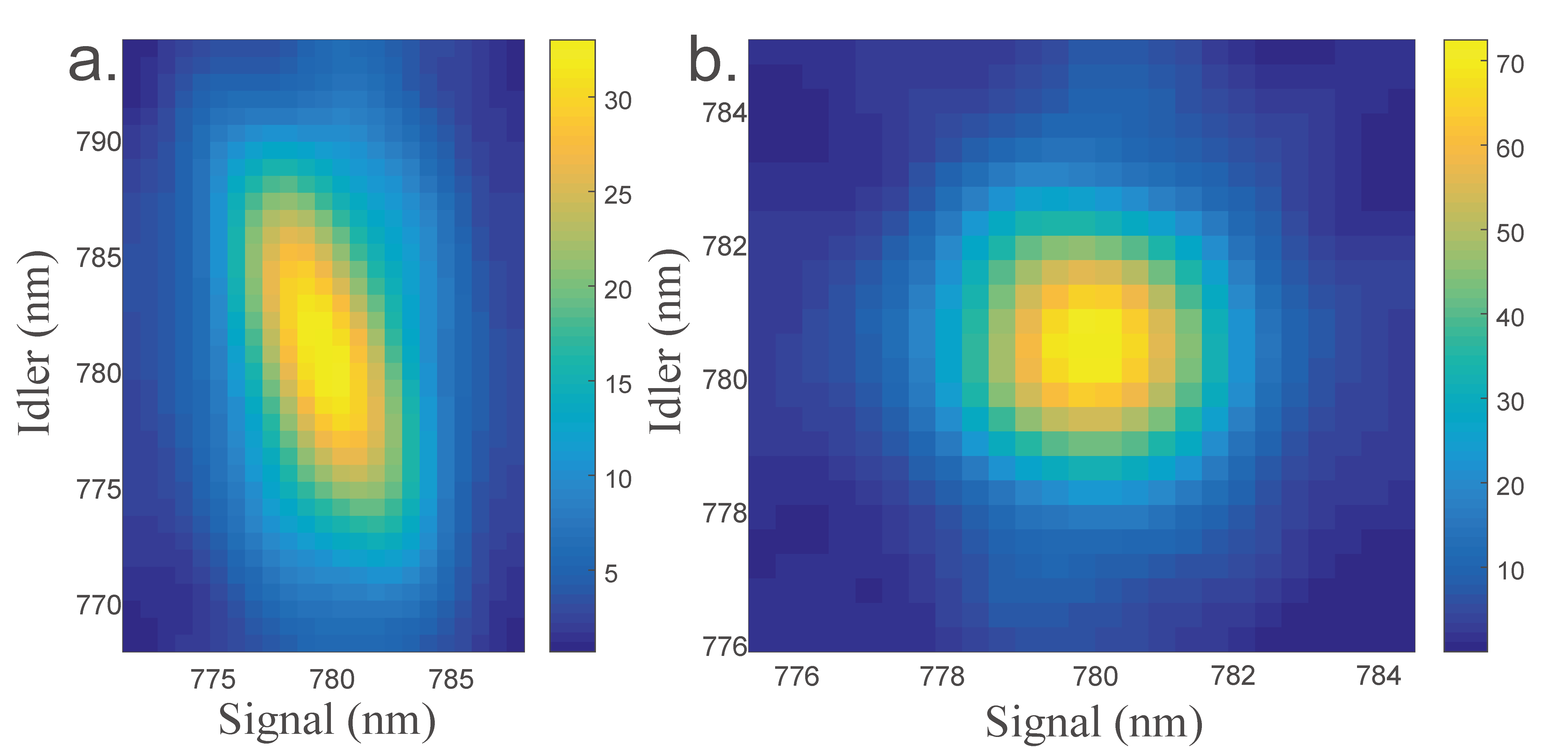}
	\caption{\textbf{Experimental results of correlation measurements with Fourier transformation}. \textbf{a} and \textbf{b} show the corrected joint spectra of correlated photons without and with bandpass filters.} 
	\label{FFT}
\end{figure}

Experimental results of time-dependent correlation matrix are obtained from $10^7$ frames (see Figure.\ref{Figure2}\textbf{c} and \textbf{d}). The rate of effective correlation events is about $10^{-3}$, in which the intrinsically low excitation rate (20\%) is an improvable factor. If applying heralded entanglement source, the rate of effective correlation events will increase substantially\cite{Pan-heralded} \cite{Zeil-heralded}. There is a tradeoff between resolution and count rate, and we choose 10 pixels as the interval between two components, which provide a resolution of $0.74 nm$ according to the calibrated spectrum distribution. 

To demonstrate COSPLI with a different joint spectrum, we use bandpass filters to release the frequency correlation, which is also an important operation to produce high-purity single photons \cite{mosley2008heralded}. The transmission spectrum of $3nm$ filters centered at $779.5 nm$ is shown in Figure.\ref{3nm}\textbf{a}. The two filters lead to a scissors matrix defined as $G(\omega_s, \omega_i) = g_s(\omega_s)g_i(\omega_i)$, where $g_s(\omega_s)$ and $g_i(\omega_i)$ are the transmission probabilities of signal and idler photons respectively. By multiplying the theoretical correlation matrix $S_{dep}(\omega_s,\omega_i)$ by $G(\omega_s, \omega_i)$ term by term, we can derive the modified theoretical correlation matrix $S_{dep}'(\omega_s,\omega_i)$ (see Figure.\ref{3nm}\textbf{b}). The measured joint spectrum is shown in Figure.\ref{3nm}\textbf{c}. In this measurement, the interval between two frequency components is changed to 5 pixels, which means that the resolution becomes $0.37 nm$. 
 
We can see that, as is shown in Figure.\ref{Figure2}\textbf{d} and Figure.\ref{3nm}\textbf{c}, the frequency components of signal and idler photons look to be discrete rather than to be continuous. It is because that we divide the continuous pattern into many segments, leading to the discontinuity of spectrum for both signal and idler photons. In order to restore it, a Fourier filter method is applied to the derived raw data. Since we have discarded all the useless information in the time space, leaving just the joint spectrum, our revised result is more smooth and clear (Figure.\ref{FFT}). 

%\section{Conclusion}
In summary, we present an approach of mapping and measuring large-scale photonic correlation with single-photon imaging. We demonstrate this by measuring the joint spectrum of correlated photons with direct imaging. Unlike previous works \cite{Kim2005Measurement,Wasilewski2006Joint, Spring2013On}, we don't have to move two detectors $m$ and $n$ times respectively to obtain a correlation matrix of $m\times n$. Other than spectrum correlation, correlations on other spaces can also be detected after being mapped to position space. The ability to address large-scale correlations in two dimensions may boost the computational power of analog quantum computing by implementing Boson sampling \cite{Aaronson2011The,Broome2013Photonic,Spring2013Boson} and quantum walk \cite{Tangeaat3174} in very large spaces.

Besides quantum technologies, COSPLI may find applications in exploring some new region of ultra-weak signals\cite{Bio-review}, like agriculture \cite{Bio-agriculture}, food chemistry \cite{Bio-food} and biomedicine \cite{Bio-biomedicine}. For example, it has been reported that neuronal activity has correlation with ultra-weak photon emission (UPE). Commercial detectors have been able to observe the ultra-weak photon emission from a cultured hippocampal slice. However, those researches are limited within the detection of accumulation images \cite{Bio-neuronal} or long time periods correlation without broader research scopes \cite{Bio-human skin}. COSPLI may serve as a entirely new technique to reveal the phenomena that correlated events happen simultaneously by short-time-UPE correlation detection.

\section*{Methods}
\noindent \textbf{Details of correlation analysis:} We first detect an accumulated image of the photons generated by spontaneous parametric down conversion (SPDC), which is shown in Figure.\ref{setup}. The wavelength range of photon pairs is very small compared to the central wavelength. Therefore, we can assume the wavelength of photons has a linear relationship with the position of them. A MATLAB program is used to divide these two beams into many parts with the same position interval. Each part represents a spectral component. After that, we detect millions of frames of correlation photon pairs by single photon ICCD. During this detecting process, electrical pulses produced by the oscillator are used to trigger ICCD so as to raise the heralding efficiency. 

Sequentially, we use a program to process the image data. The first step is to calculate the average background luminance ($I_b$) of the image. Based on the background we set a gap ($I_g$), and every spot possessing a luminance above $I_b + I_g$ will be considered as a photon. Under this condition, this program records the position of each recorded photon. Correlation data is recorded if and only if two photons appear in the signal and idler area, respectively. In the data accumulating process, efficiency and losses are significant. Among these factors, heralding efficiency of oscillator trigger is determained by the excitation rate, which is about 20\%. Setup transmission, mainly influenced by the diffraction loss, is 50\%. And detecting efficiency of ICCD is 20\% at the wavelength of $780nm$, which should be squared due to two photons. Because of these, about 1000 frames contain one correlation data.\\

\noindent \textbf{Minimize noises:}
To ensure high visibility of photons, we erase any possible stray light with respect to the spectral, temporal and spatial domains. A $12nm$ band pass filter centered at $780nm$ is applied before the ICCD to reduce the stray light from the environment. In the time domain, electric pulses produced by the oscillator are used to trigger ICCD in order to ensure high heralding efficiency. Particularly, the travel time of electrical trigger and photons are not necessary to equal. Because the gate width nearly equals to the laser pulse interval, there must be no more than one pair of photons in every detect window of ICCD, even though the photons detected may not be generated by the pulse sending the trigger. At last, in free space, we use light-blocking materials to cover every part of the optical setup.\\

\noindent \textbf{Joint spectrum:}
In frequency space, the joint spectrum intensity should be written as $S_{dep}(\omega_s,\omega_i) = \left |f(\omega_s,\omega_i)\right |^2$, where {\it f} is the probability amplitude of two photon state generated by SPDC.
\begin{equation}
f(\omega_s,\omega_i) = A\alpha(\omega_s+\omega_i)\phi(\omega_s,\omega_i),
\end{equation} 
In this equation, {\it A} is a normalization constant, $\alpha(\omega_s+\omega_i)$ is a function describing the envelope of the pump field. Due to the conservation of energy, we assume the envelope is a Gussian-like beam centered at the line $\omega_s + \omega_i = const$, the width of which is determined by the spectral width of the pump laser. $\phi(\omega_s,\omega_i)$ is a phase matching function given by
\begin{equation}
\begin{split}
\phi(\omega_s,&\omega_i)  = sinc \{\frac{L}{2}[k_p(\omega_s+\omega_i) - k_s(\omega_s)-k_i(\omega_i)]\}\\
&\times\exp\{-i\frac{L}{2}[k_p(\omega_s+\omega_i) - k_s(\omega_s)-k_i(\omega_i)]\}
\end{split}	
\end{equation}

\noindent \textbf{Two-dimensional Fourier filter:} If we calculate the Fourier transform of raw data, there will exist a central direct current (DC) peak, a vertical line, a horzontal line and some noise. The central DC peak contains the information of the the real joint spectrum, while those two lines are generated from the data collecting process, which should be abandoned. Therefore, the continuous joint spectrum will be restored by just doing inverse Fourier transform on the DC peak only. 

\section*{Acknowledgements} 
The authors thank Jian-Wei Pan for helpful discussions. This research is supported by the National Key Research and Development Program of China (2017YFA0303700), National Natural Science Foundation of China (Grant No. 61734005, 11761141014, 11690033, 11374211), the Innovation Program of Shanghai Municipal Education Commission, Shanghai Science and Technology Development Funds, and the open fund from HPCL (No. 201511-01), X.-M.J. acknowledges support from the National Young 1000 Talents Plan.\\

%\newpage

\end{document}